\documentclass[pra,twocolumn,superscriptaddress,showpacs,floatfix]{revtex4}

\usepackage{epsfig}
\usepackage{epsf}
\usepackage{bm}                 
\usepackage{amsmath}
\usepackage{amssymb}
\usepackage{dcolumn}
\usepackage{graphicx}
\usepackage{psfrag}
\usepackage{latexsym}

\def\unity{\mbox{\small 1} \!\! \mbox{1}}

\newcommand{\ket}[1]{|#1\rangle}
\newcommand{\bra}[1]{\langle#1|}

\begin{document}
\title{Repeat-Until-Success quantum computing using stationary and
  flying qubits}

\author{Yuan Liang Lim}
\affiliation{Blackett Laboratory, Imperial College London, Prince
  Consort Road, London SW7 2BZ, United Kingdom}

\author{Sean D. Barrett}
\affiliation{Hewlett-Packard Laboratories, Filton Road, Stoke Gifford,
  Bristol BS34 8QZ, United Kingdom}

\author{Almut Beige}
\affiliation{Blackett Laboratory, Imperial College London, Prince
  Consort Road, London SW7 2BZ, United Kingdom}

\author{Pieter Kok}
\affiliation{Hewlett-Packard Laboratories, Filton Road, Stoke Gifford,
  Bristol BS34 8QZ, United Kingdom}

\author{Leong Chuan Kwek}
\affiliation{National Institute of Education, Nanyang Technological
  University, Singapore 63 9798, Singapore}
\affiliation{Department of Physics, National University of Singapore,
  Singapore 11 7542, Singapore}

\date{\today}

\begin{abstract}
 We introduce an architecture for robust and scalable quantum
 computation using both stationary qubits (e.g.~single photon sources
 made out of trapped atoms, molecules, ions, quantum dots, or defect
 centers in solids) and flying qubits (e.g.~photons). Our scheme
 solves some of the most pressing problems in existing non-hybrid
 proposals, which include the difficulty of scaling conventional
 stationary qubit approaches, and the lack of practical means for
 storing single photons in linear optics setups. We combine elements
 of two previous proposals for distributed quantum computing, namely
 the efficient photon-loss tolerant build up of cluster states by
 Barrett and Kok [Phys. Rev. A {\bf 71},   060310 (2005)] with the
 idea of Repeat-Until-Success (RUS) quantum computing by Lim {\em et
 al.} [Phys. Rev. Lett. {\bf 95}, 030505   (2005)]. This idea can be
 used to perform eventually deterministic two-qubit logic gates on
 spatially separated stationary qubits via photon pair
 measurements. Under non-ideal conditions, where photon loss is a
 possibility, the resulting gates can still be used to build graph
 states for one-way quantum computing. In this paper, we describe the
 RUS method, present possible experimental realizations, and analyse
 the generation of graph states. 
\end{abstract}


\maketitle

\section{Introduction}

\noindent
Quantum computing offers a way to realize certain algorithms
exponentially more efficiently than with the best known classical
solutions \cite{shor,deutsch}. A substantial effort has therefore been
made to develop the corresponding quantum
technologies. Proof-of-principle experiments demonstrating the
feasibility of quantum computing have already been performed: Using
nuclear magnetic resonance techniques, Vandersypen {\em et al.}
\cite{chuang} realized a simple instance of Shor's algorithm by
factoring $15 = 3 \times 5$. A two qubit gate has been implemented in
a color center in diamond, utilizing the electron spin state of the
nitrogen-vacancy defect center together with a nearby nuclear spin as
qubits \cite{jelezko2004b}. Groups in Innsbruck and Boulder
implemented a universal two-qubit gate in an ion trap
\cite{blatt,wineland}, and the three-qubit teleportation protocol
\cite{blatt2,wineland2}. Adding more qubits to this ``proto quantum
computer'' will increase the density of the motional states used for
the two-qubit interaction. Consequently, it will become even harder to
implement clean two-qubit gates. Scaling ion trap quantum computers
much further therefore seems to require some form of distributed
quantum information processing, possibly involving ion transport
\cite{kielpinski}.

The schemes mentioned above are based on manipulating {\em stationary}
qubits such as atoms, molecules, or trapped ions. An alternative route
to finding a feasible and scalable technology for building quantum
computers is based on {\em flying} qubits, such as photons. The main
advantage of photons is their extremely long coherence time: In vacuum
and in simple dielectric media, photons do not interact with their
environment, and hence do not lose their quantum information. This is
why photons are usually the qubits of choice for quantum communication
\cite{bennett,ekert}. However, at the same time this lack of
interaction means that it is very hard to create two-photon entangling
gates. It therefore came as a surprise that the bosonic symmetry
requirement of the electromagnetic field, together with photon
counting and proper single-photon sources, is sufficient for
implementing scalable quantum computing \cite{KLM}. The overhead cost
for Linear Optical Quantum Computing (LOQC) has subsequently been
brought down significantly. In particular, the one-way or
cluster-state model for quantum computing \cite{raussendorf} has
allowed for drastic improvements in the scalability
\cite{yoran03,nielsen04,browne04}. Recently, a four-qubit cluster
state was realized experimentally by Walther {\em et al}.\
\cite{walther}. The main drawbacks of LOQC are the difficulties of
maintaining interferometric stability, the lack of practical `on
demand' single-photon sources and the lack of quantum memories for
photonic qubits \cite{gingrich03}.

\begin{figure}
\begin{minipage}{\columnwidth}
\begin{center}
\resizebox{\columnwidth}{!}{\rotatebox{0}{\includegraphics{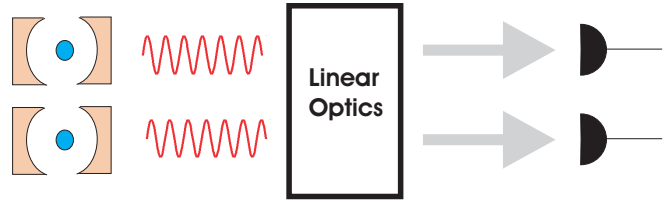}}}
\end{center}
\vspace*{-0.5cm}
\caption{Experimental realization of a universal two-qubit gate for
  the considered network of single photon sources (stationary
  qubits). This requires the generation of a photon
  within each of the sources involved. The two photons then pass within
  their coherence time through a linear optics network, which performs
  a certain photon pair measurement.} \label{moon}
\end{minipage}
\end{figure}

In this paper, we consider the practical advantage of combining
stationary and flying qubits for the realisation of scalable quantum
computing. The stationary qubits (single photon sources) are arranged
in a network of nodes with each node processing and storing a small
number of qubits. To achieve scalability, the concept of distributed
quantum computing was introduced and it was proposed that distant
qubits communicate with each other through the means of flying qubits
(i.e.~photons) \cite{Grover,Cirac}. Initial schemes for the
implementation of this idea relied on entangled ancillas as a resource
\cite{Grover,Cirac,Jens,duan04,taylor04}. Others required that the
photon from one source is fed into another source
\cite{cirac97,Enk,Molmer,meier04,Guo,Zhou} or a photon-mediated
interaction between two fiber-coupled distant cavities needed to be
established \cite{Bose}. More hybrid approaches to quantum computing 
can be found in Refs.~\cite{franson,bill}.

Other authors developed schemes for the probabilistic generation of
highly entangled states between distant single photon sources
\cite{cabrillo,bose99,zoller,last,browne,Simon,lim,Zou,chim}. In these
schemes, one generates a photon in each of the sources and then
performs an entangling photon measurement. By virtue of entanglement
swapping, this results in entangled stationary qubits. It has been
shown that similar ideas can also result in the implementation of
probabilistic remote two-qubit gates \cite{grangier}. At this point,
it was believed that scalable quantum computing with distant photon
sources requires additional resources such as local entangling gates
\cite{duan04} or entangled ancillas in order to become deterministic.

The concrete setup that we consider in this paper has recently been
introduced by Lim {\em et al.}  \cite{moonlight} and allows for the
more efficient implementation of universal two-qubit gates than
previous proposals. The presented scheme 
consists of a network of single stationary qubits (like trapped atoms,
molecules, ions, quantum dots and nitrogen vacancy color centers)
inside optical cavities, which act as a source for the generation of
single photons on demand. Read-out measurements and single qubit
rotations can be performed on the stationary qubits using laser pulses
and standard quantum optics techniques as employed in the recent ion
trap experiments in Innsbruck and Boulder \cite{blatt,wineland}.

The main building block for the realization of a two-qubit gate, which
qualifies the setup for universal quantum computing, is shown in
Figure \ref{moon}. It requires the simultaneous generation of a photon
in each source involved in the operation. Afterwards the photons
should pass through a linear optics setup, where a pair measurement is
performed in the output ports. This photon pair measurement results
either in the completion of the gate or indicates the presence of the
original qubits. In the later event, the gate should be repeated. The
qubits are never lost in the computation and the presented scheme has
therefore been called {\em Repeat-Until-Success} quantum
computing \cite{moonlight}.

Under realistic conditions, i.e.~in the presence of finite
detector efficiencies and finite success rates for the
generation of a single photon on demand, the setup in
Figure 1 can still be used for the implementation of
probabilistic gates with a very high fidelity. As shown
recently by Barrett and Kok \cite{Sean}, it is possible to use
probabilistic gates to efficiently generate graph states
for one-way quantum computing \cite{raussendorf}. Both
schemes, \cite{moonlight} and \cite{Sean} overcome the
limitations to scalable quantum computing faced before when
using the same resources. In Ref.~\cite{moonlight} this is
achieved with an eventually deterministic gate.
Ref.~\cite{Sean} introduced a so-called double-heralding
scheme, in which the entangling photo-detection stage was
employed twice to eliminate unwanted separable
contributions to the density matrix.

In this paper, we combine the ideas presented in our previous work
\cite{moonlight,Sean}. In this way, we obtain a truly scalable design
for quantum computing, i.e.~even in the presence of imperfect
components, with several key advantages:
\begin{enumerate}
 \item Since our system uses {\em no direct qubit-qubit interactions},
   the qubits can be well isolated. Not only does this allow us to
   address the individual qubits easily, it also means that there are
   fewer decoherence channels and hence fewer errors in the computation.
\item We achieve {\em robustness to photon loss}. In the
presence of photon loss, the two qubit gates become
non-deterministic.  However, the gate failures are
heralded, and so the gates can still be used to build
high-fidelity entangled states, albeit in a
non-deterministic manner. Photon loss thus increases the
overall overhead cost associated with the scheme, but does
not directly reduce the fidelity of the computation. When
realistic photo-detectors and optical elements are used,
photon loss is inevitable and this built-in robustness is
essential.
 \item Our scheme largely relies only on {\em components that have
   been demonstrated in experiments} like atom-photon entanglement
   \cite{Monroe,weber}. Apart from linear optics, we require only
   relatively good sources for the generation of single photons on
   demand \cite{Kuhn2,Yamamoto,Mckeever,Lange04,Kuzmich}, preferrably
   at a high rate \cite{grangierscience}, and relatively efficient but
   not necessarily number resolving photon detectors
   \cite{Rosenberg}. Combining these in a working 
   quantum computer will be challenging, but the basic physics has
   been shown to be correct. 
 \item The photon pair measurement is {\em interferometrically
 stable}. Since each generated photon contributes equally to the
 detection of a photon in the linear optics setup, fluctuations in the
 length between the photon source and the detectors can at most result
 in an overall phase factor with no physical consequences. This
 constitutes a significant advantage compared to previous schemes based on
 one-photon measurements (the only interferometrically stable schemes
 are \cite{last,Simon,lim,moonlight}), since the
 photons do not need to arrive 
 simultaneously in the detectors as long as they overlap within their
 coherence time in the setup.
 \item The basic ideas presented in this paper are {\em
 implementation independent} and the stationary qubits can be realised
 in a variety of ways. Any system with the right energy-level
 structure and able to produce encoded flying qubits may be used. 
 \item Our scheme is inherently {\em distributed}. Hence, it can be
   used in applications which integrate both quantum
   computation and quantum communication. We show that entanglement
   can be generated directly between any two 
   stationary qubits in the physical quantum computer. This
   significantly reduces the computational cost compared to
   architectures involving only nearest-neighbor interactions between
   the qubits \cite{gottesman2000}.
\end{enumerate}

This paper is organized as follows. In the next Section, we give an
overview on the basic principles of measurement-based quantum
computing, since the described hybrid approach to quantum computing
constitutes a novel implementation of these ideas. Section III details
the general principle of a remote two-qubit gate implementation. In
Section \ref{rea}, we discuss possible gate implementations with
polarization and time-bin encoded photons. In Section \ref{cluster},
we describe how to overcome imperfections of inefficient photon
generation and detection with the help of 
pre-fabricated graph states. Finally, we state our conclusions in
Section \ref{conc}.

\section{Measurement-based quantum computing} \label{sectionii} 

\noindent
One condition for the successful implementation of a measurement-based
quantum gate is that the measurement outcome is {\em mutually
  unbiased} \cite{Wootters} with respect to the computational
basis. In this way, an observer does not learn anything about the
state of the qubits and the information might remain stored inside the
computer. To avoid the destruction of qubits, it is not allowed to
measure on the qubits directly. Measurements should only be performed
on ancillas, which have interacted and therefore share entanglement
with the qubits. These ancillas can be of the same physical
realisation as the computational qubits
\cite{nine,fifteen,raussendorf,leung} but they might also be realised
differently. If the stationary qubits are atoms, the ancilla can be
the quantised field mode inside an optical cavity \cite{beige}, a
common vibrational mode \cite{beige2} or newly generated photons, as
in the setup considered here. Vice versa, it has been found
advantageous to use collective atomic states as ancillas for photonic
qubits \cite{franson,bill}.  

Let us now briefly describe the principles of measurement-based
quantum computing in a more formal way. Using the terms `qubits' and
`ancillas' provides a convenient picture, which is especially suited
for the description of hybrid approaches, where the qubits may remain
encoded in the same physical qubits instead of being assigned
dynamically as the computation proceeds. As in Ref.~\cite{kok}, we
consider two systems, $s$ and $a$, that are initially in the state
($c_n \in \mathbb{C}$) 
\begin{equation}
 |\Psi\rangle_s |A_0\rangle_a \equiv \sum_n c_n |\psi_n\rangle_s \otimes
 |A_0\rangle_a \, .
\end{equation}
After some interaction, the joint system evolves into
\begin{equation}
 |\Psi\rangle_s |A_0\rangle_a \rightarrow \sum_n c_n |\psi_n\rangle_s
 \otimes |A_n\rangle_a \equiv |\Phi\rangle \, ,
\end{equation}
where the $|A_n\rangle_a$ are the eigenstates of an observable
$\mathsf{A}$. We can then measure $\mathsf{A}$, which
will reveal the state of the system $s$. This can be interpreted as a
quantum non-demolition measurement of $s$ but this is not what we are
interested in here. 

\begin{figure}[t]
   \begin{center}
   \begin{psfrags}
     \psfrag{in}[r]{$|\Psi\rangle_s$}
     \psfrag{out}{$|\Upsilon_k\rangle_s$}
     \psfrag{aux}[r]{$|A_0\rangle_a$}
     \psfrag{m}{$\hat{B}_k$}
     \psfrag{U}{$U_{sf}$}
        \epsfig{file=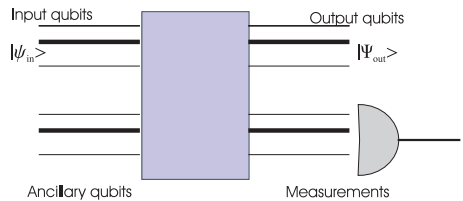, width=6cm}
   \end{psfrags}
   \end{center}
  \caption{Measurement-based quantum computing. The input state
  $|\Psi\rangle_s$ and the auxiliary state $|A_0\rangle_a$ are transformed
  in an $N$-port that induces a unitary transformation $U_{sf}$. Given a
  detector outcome corresponding to a POVM $\hat{B}_k$, the output
  state is $|\Upsilon_k\rangle_s$.} \label{minkslevel}
\end{figure}

In this article we will instead consider measurements of an
observable $\mathsf{B}$, as shown in Figure \ref{minkslevel}, that is
complementary to $\mathsf{A}$. In other words, the eigenvectors of
$\mathsf{A}$ and $\mathsf{B}$ form a so-called mutually unbiased basis
of the Hilbert space of system $a$. A specific outcome labelled $k$ of
such a measurement corresponds to the application of the projection
operator $\hat{B}_k$ (associated with the $k^{\mathrm{th}}$
eigenvector of $\mathsf{B}$), and the state of system $s$ is then
given by 
\begin{equation}
 |\Upsilon_k\rangle_s = \frac{ \text{Tr}_a \left[ \langle\Phi| \unity
  \otimes \hat{B}_k |\Phi\rangle \right]}{\text{Tr}_{sa} \left[
  \langle\Phi| \unity \otimes \hat{B}_k |\Phi\rangle \right]} \, .
\end{equation}
This can be generalized to situations where $\hat{B}_k$ is a
multi-rank projector or a Positive Operator Valued Measure (POVM). The
conditions for the evolution $|\psi_n\rangle_s \to |\Upsilon_k
\rangle_s$ to be a unitary transformation on system $s$ are presented
in Lapaire {\em et al.} \cite{kok}. If $s$ describes the qubits and
$a$ the ancilla, they guarantee, as mentioned above, that the
detection of $\hat{B}_k$ does not reveal any information about the
qubits. 

Especially, in the setup considered in this paper the system $s$
consists of a set of $N$ stationary qubits occupying a Hilbert space
of size $2^N$, and system $a$ consists of $N$ flying quantum systems
occupying a Hilbert space of dimension $d \geq 2^N$. A measurement of
the observable $\mathsf{B}$ on the flying qubits will result in a
multi-qubit (entangling) operation on the stationary qubits. We are
interested in the case where the projector $\hat{B}_k$ induces a {\em
  unitary} transformation on the stationary qubits,
\begin{equation}
 |\Upsilon_k\rangle_s = U_k |\Psi\rangle_s \, ,
\end{equation}
which means that $\hat{B}_k$ is a proper projector. There are two
interesting cases to consider:
\begin{enumerate}
  \item The set $\{\hat{B}_k\}_a$ corresponds to a basis of states
   that induces a complete set $\{ U_k\}_a$ of entangling quantum
   gates. As a result, finding any measurement outcome $k$ will induce
   a unitary entangling gate operation on the stationary qubits. 
  \item The set $\{\hat{B}_k\}_a$ corresponds to a basis that can be
   divided into two sets of states: Some of the projectors will induce
   a unitary entangling gate $U_k$ on the stationary qubits, while the
   remaining projectors induce a transformation that is locally
   equivalent to the identity map $\unity$.
\end{enumerate}

The second setup is interesting for the following reason: Suppose that
system $s$ consists of $N$ non-interacting (e.g., well-separated)
stationary qubits with long decoherence times. If this system can
generate flying qubits in the manner described above, we can perform a
measurement of the observable $\mathsf{B}$ and entangle the
non-interacting stationary qubits. When not all measurement outcomes
produce an entangling gate on the stationary qubits (some yield
instead the identity operator), then the unitary gate is applied only 
part of the time. However, due to the fact that a gate failure corresponds to
an identity operation (or something locally equivalent), we can again
prepare the flying qubits in the state $|A_0\rangle_a$. This allows us
to repeat the protocol 
until the entangling gate is  applied successfully, which is why we
called this idea Repeat-Until-Success quantum computing \cite{moonlight}. 

\section{Remote two-qubit phase gates}

\noindent
One of the requirements for universal quantum computing is the ability
to perform an entangling two-qubit gate operation, like a controlled 
phase gate \cite{divincenzo}. In this Section we describe the general
concept for the implementation of this gate between two distant single
photon sources. Note that our method of  distributed quantum computing
allows to realize entangling operations, since the measurement on a
photon pair can imprint a phase on the state of its sources although
it cannot change the distribution of their populations. The first step
for the implementation of a two-qubit gate is the generation of a
photon within each respective source, which encodes the information of
the stationary qubit. 

\subsection{Encoding}

\noindent
Let us denote the states of the photon sources, which encode the
logical qubits $|0 \rangle_{\rm L}$ and $|1 \rangle_{\rm L}$ as $|0
\rangle$ and $|1 \rangle$, respectively.  An arbitrary pure state of
two stationary qubits can then be written as
\begin{equation} \label{original}
\ket{\psi_{\rm in}}=\alpha \, \ket{00} + \beta\, \ket{01} + \gamma \,
\ket{10} + \delta \, \ket{11} \, ,
\end{equation}
where $\alpha$, $\beta$, $\gamma$ and $\delta$ are 
complex coefficients with $|\alpha|^2 + |\beta|^2 + |\gamma|^2 +
|\delta|^2=1$. Suppose that a photon is generated in each of the two
sources, whose state (i.e.~its polarization  or generation
time) depends on the state of the respective source. In
the following we assume that source 1 prepared in $|i \rangle$ leads
to the creation of one photon in state $|{\sf x}_i \rangle$, while
source 2  prepared in $|i \rangle$ leads to the creation of one photon
in state $|{\sf y}_i \rangle$, 
\begin{equation} \label{enc}
\ket{i}_1 \rightarrow \ket{i ,{\sf x}_i}_1 \, , ~~ \ket{i}_2
\rightarrow \ket{i ,{\sf y}_i}_2 \, .
\end{equation}
The simultaneous creation of a photon in both sources then transfers
the initial state (\ref{original}) into
\begin{eqnarray} \label{theencoding}
\ket{\psi_{\rm enc}} &=& \alpha \, \ket{00 ,{\sf x}_0{\sf y}_0} + \beta
\, \ket{01 ,{\sf x}_0{\sf y}_1} + \gamma \,   \ket{10 ,{\sf x}_1{\sf
y}_0} \nonumber \\ && +\delta \, \ket{11 ,{\sf x}_1{\sf y}_1} \, .
\end{eqnarray}
Note that the generation of photons whose state depends on the states
of the stationary qubits is a highly non-linear process. The
preparation of the generally entangled state (\ref{theencoding}) is
indeed the key step which allows the completion of an
eventually deterministic two-qubit gate with otherwise nothing else
than linear optics and photon pair measurements. The way the encoding step
(\ref{enc}) can be realized experimentally is discussed in Section
\ref{rea}. In this section we focus on the general ideas underlying
Repeat-Until-Success quantum computing.

\subsection{Mutually Unbiased Basis}

\noindent
Once the photons have been created, an entangling phase gate can be
implemented by performing an absorbing measurement on the photon
pair. Thereby, it is important to choose the photon measurement such
that none of the possible outcomes reveals any information about the
coefficients $\alpha$, $\beta$, $\gamma$ and $\delta$, as mentioned in
Section II. This can be achieved with a
photon pair measurements in a basis mutually unbiased  \cite{Wootters} with respect to
the computational basis given by the states $\{\ket{{\sf x}_0{\sf
y}_0}, \, \ket{{\sf x}_0{\sf y}_1}, \, \ket{{\sf x}_1{\sf y}_0}, \,
\ket{{\sf x}_1{\sf y}_1} \}$. More concretely, all possible outcomes
of the photon measurement should be of the form
\begin{equation} \label{unbiased}
\ket{\Phi} = {\textstyle {1 \over 2}} \big[ \ket{{\sf x}_0{\sf y}_0}
  + {\rm e}^{{\rm i} \varphi_1} \, \ket{{\sf x}_0{\sf y}_1} + {\rm
  e}^{{\rm i} \varphi_2} \, \ket{{\sf x}_1{\sf y}_0} + {\rm e}^{{\rm
  i} \varphi_3} \, \ket{{\sf x}_1{\sf y}_1} \big] \, . 
\end{equation}
As we see below, a complete set of basis states of this form can be
found. Any bias in the
amplitudes would yield information about $\alpha$, $\beta$, $\gamma$,
and $\delta$, and would therefore not induce a unitary gate on the
stationary qubits. Detecting the state (\ref{unbiased}) and absorbing
the two photons in the process transfers the encoded state
(\ref{theencoding}) into 
\begin{eqnarray} \label{out}
\ket{\psi_{\rm out}} &=& \alpha \, \ket{00} + {\rm e}^{-{\rm i}
  \varphi_1} \, \beta \, \ket{01} + {\rm e}^{-{\rm i} \varphi_2} \,
  \gamma \, \ket{10} \nonumber \\ && + {\rm e}^{-{\rm i} \varphi_3} \,
  \delta \, \ket{11} \, .
\end{eqnarray}
Note that the output state (\ref{out}) differs from the initial state
(\ref{original}) by a two-qubit phase gate.

Let us now consider the angle
\begin{equation}
\varphi_3 = \varphi_1 + \varphi_2 \, .
\end{equation}
In this case, the state $\ket{\Phi}$ is a product state and the output
(\ref{out}) differs from the initial state (\ref{original}) only by
local operations. However, if
\begin{equation} \label{phasecondition}
\varphi_3 = \varphi_1 + \varphi_2 + \pi \, ,
\end{equation}
the state (\ref{unbiased}) becomes a maximally entangled state, as it
becomes obvious when writing $\ket{\Phi}$ as
\begin{equation}
|\Phi \rangle =  {\textstyle {1 \over 2}} \big[ \ket{{\sf x}_0} (
  \ket{{\sf y}_0} + {\rm e}^{{\rm i} \varphi_1} \, \ket{{\sf y}_1} )+
      {\rm e}^{{\rm i} \varphi_2} \, \ket{{\sf x}_1} (\ket{{\sf y}_0}
      -   {\rm e}^{{\rm i} \varphi_1} \, \ket{{\sf y}_1}) \big] \, .
\end{equation} 
The detection of a photon pair in this maximally entangled state
results in the completion of a phase gate with maximum entangling
power on the stationary qubit. Vice versa, maximum entanglement of the
state (\ref{unbiased}) also automatically implies
Eq.~(\ref{phasecondition}) as one can show by calculating the
entanglement of formation of the state (\ref{unbiased}).  

\subsection{A deterministic gate}

\noindent
In the following, we denote the states of the measurement basis,
i.e. the mutually unbiased basis, by $\{ \ket{\Phi_i} \}$. In order to 
find a complete Bell basis with all states of form (\ref{unbiased}),
we define
\begin{eqnarray} \label{base}
\ket{\Phi_1} &\equiv & {\textstyle {1 \over \sqrt{2}}} \big[ \ket{{\sf
      a}_1 {\sf b}_1}+\ket{{\sf a}_2 {\sf b}_2} \big] \, , \nonumber
      \\ \ket{\Phi_2} &\equiv & {\textstyle {1 \over \sqrt{2}}} \big[
      \ket{{\sf a}_1 {\sf b}_1}-\ket{{\sf a}_2 {\sf b}_2} \big] \, ,
      \nonumber \\ 
      \ket{\Phi_3} &\equiv & {\textstyle {1 \over
      \sqrt{2}}} \big[ \ket{{\sf a}_1 {\sf b}_2}+\ket{{\sf a}_2 {\sf
      b}_1} \big] \, , \nonumber \\ \ket{\Phi_4} &\equiv & {\textstyle
      {1 \over \sqrt{2}}} \big[ \ket{{\sf a}_1 {\sf b}_2}-\ket{{\sf
      a}_2 {\sf b}_1} \big] \, ,
\end{eqnarray}
where the states $|{\sf a}_i \rangle$ describe photon 1 and the $|{\sf
  b}_i \rangle$ describe photon 2. Assuming orthogonality,
  i.e.~$\langle {\sf a}_1 | {\sf a}_2 \rangle =0$ and $\langle {\sf
  b}_1 | {\sf b}_2 \rangle = 0$, one can write the photon states on
  the right hand side of Eq.~(\ref{base}) without loss of generality as
\begin{eqnarray} \label{ab}
\ket{{\sf a}_1} &=& c_1 \, \ket{{\sf x}_0} +  {\rm e}^{{\rm i}
  \vartheta_1} s_1 \, \ket{{\sf x}_1} \, , \nonumber \\ \ket{{\sf
  a}_2} &=& {\rm e}^{-{\rm i} \xi_1}(  {\rm e}^{-{\rm i} \vartheta_1}
  s_1\, \ket{{\sf x}_0} -  c_1 \, \ket{{\sf x}_1}) \, ,\nonumber \\
  \ket{{\sf b}_1} &=& c_2 \, \ket{{\sf y}_0} + {\rm e}^{{\rm i}
  \vartheta_2} s_2 \,\ket{{\sf y}_1} \, , \nonumber \\ \ket{{\sf b}_2}
  &=& {\rm e}^{-{\rm i} \xi_2}({\rm e}^{-{\rm i} \vartheta_2} s_2 \,
  \ket{{\sf y}_0} - c_2 \, \ket{{\sf y}_1})
\end{eqnarray}
with
\begin{eqnarray}
s_i \equiv \sin{\theta_i} \, , ~~ c_i \equiv \cos{\theta_i} \, .
\end{eqnarray}
Inserting this into Eq.~(\ref{base}), we find
\begin{widetext}
\begin{eqnarray} \label{base2}
\ket{\Phi_1} &=& {\textstyle {1 \over \sqrt{2}}} \big[ \big
  (c_1c_2+{\rm e}^{-{\rm i}(\vartheta_1+\vartheta_2)} {\rm e}^{-{\rm
  i}(\xi_1+\xi_2)} s_1s_2 \big) \ket{{\sf x}_0{\sf y}_0} + \big( {\rm
  e}^{{\rm i}\vartheta_2}c_1s_2 - {\rm e}^{-{\rm i}\vartheta_1} {\rm
  e}^{-{\rm i}(\xi_1+\xi_2)} s_1c_2 \big)   \ket{{\sf x}_0{\sf y}_1}
  \nonumber \\ &&+ \big( {\rm e}^{{\rm i}\vartheta_1}s_1c_2-{\rm
  e}^{-{\rm i}\vartheta_2} {\rm e}^{-{\rm i}(\xi_1+\xi_2)} c_1s_2
  \big) \ket{{\sf x}_1{\sf y}_0}+ \big( {\rm e}^{{\rm
  i}(\vartheta_1+\vartheta_2)}s_1s_2+ {\rm e}^{-{\rm i}(\xi_1+\xi_2)}
  c_1c_2 \big) \ket{{\sf x}_1{\sf   y}_1} \big],  \nonumber \\
  \ket{\Phi_2} &=& {\textstyle {1 \over \sqrt{2}}} \big[
  \big(c_1c_2-{\rm e}^{-{\rm i}(\vartheta_1+\vartheta_2)} {\rm
  e}^{-{\rm i}(\xi_1+\xi_2)} s_1s_2 \big)\ket{{\sf x}_0{\sf y}_0} +
  \big({\rm e}^{{\rm i}\vartheta_2}c_1s_2+{\rm e}^{-{\rm
  i}\vartheta_1} {\rm e}^{-{\rm i}(\xi_1+\xi_2)} s_1c_2 \big)
  \ket{{\sf x}_0{\sf y}_1} \nonumber \\  && + \big( {\rm e}^{{\rm
  i}\vartheta_1}s_1c_2+{\rm e}^{-{\rm i}\vartheta_2} {\rm e}^{-{\rm
  i}(\xi_1+\xi_2)} c_1s_2 \big) \ket{{\sf x}_1{\sf y}_0}+ \big( {\rm
  e}^{{\rm i}(\vartheta_1+\vartheta_2)}s_1s_2-{\rm e}^{-{\rm
  i}(\xi_1+\xi_2)} c_1c_2 \big) \ket{{\sf x}_1{\sf y}_1} \big],
  \nonumber \\  \ket{\Phi_3} &=&{\textstyle {1 \over \sqrt{2}}}\big[
  \big({\rm e}^{-{\rm i}\vartheta_2} {\rm e}^{-{\rm
  i}\xi_2}c_1s_2+{\rm e}^{-{\rm i}\vartheta_1} {\rm e}^{-{\rm i}\xi_1}
  s_1c_2 \big)\ket{{\sf x}_0{\sf y}_0}- \big({\rm e}^{-{\rm i}\xi_2}
  c_1c_2-{\rm e}^{-{\rm i}(\vartheta_1-\vartheta_2)} {\rm e}^{-{\rm
  i}\xi_1}s_1s_2 \big) \ket{{\sf x}_0{\sf y}_1} \nonumber \\  &&+
  \big({\rm e}^{{\rm i}(\vartheta_1-\vartheta_2)} {\rm e}^{-{\rm
  i}\xi_2} s_1s_2- {\rm e}^{-{\rm i}\xi_1} c_1c_2 \big) \ket{{\sf
  x}_1{\sf y}_0} - \big({\rm e}^{{\rm i}\vartheta_1} {\rm e}^{-{\rm
  i}\xi_2} s_1c_2+{\rm e}^{{\rm i}\vartheta_2} {\rm e}^{-{\rm
  i}\xi_1}c_1s_2 \big) \ket{{\sf x}_1{\sf y}_1} \big] , \nonumber \\
  \ket{\Phi_4} &=&{\textstyle {1 \over \sqrt{2}}}\big[ \big({\rm
  e}^{-{\rm i}\vartheta_2} {\rm e}^{-{\rm i}\xi_2} c_1s_2-{\rm
  e}^{-{\rm i}\vartheta_1} {\rm e}^{-{\rm i}\xi_1}s_1c_2 \big)
  \ket{{\sf x}_0{\sf y}_0} - \big({\rm e}^{-{\rm i}\xi_2} c_1c_2+{\rm
  e}^{-{\rm i}(\vartheta_1-\vartheta_2)} {\rm e}^{-{\rm i}\xi_1}
  s_1s_2 \big) \ket{{\sf x}_0{\sf y}_1} \nonumber \\  && + \big({\rm
  e}^{{\rm i}(\vartheta_1-\vartheta_2)} {\rm e}^{-{\rm i}\xi_2}
  s_1s_2+ {\rm e}^{-{\rm i}\xi_1} c_1c_2 \big) \ket{{\sf x}_1{\sf
  y}_0}- \big({\rm e}^{{\rm i}\vartheta_1} {\rm e}^{-{\rm i}\xi_2}
  s_1c_2-{\rm e}^{{\rm i}\vartheta_2} {\rm e}^{-{\rm i}\xi_1} c_1s_2
  \big) \ket{{\sf x}_1{\sf y}_1} \big]  .  \nonumber \\
\end{eqnarray}
\end{widetext}
These states are of the form (\ref{unbiased}), if the amplitudes are
all of the same size, which yields the conditions
\begin{eqnarray} \label{constraint1}
&& \hspace*{-1.6cm} \big| c_1c_2\pm{\rm e}^{-{\rm
      i}(\vartheta_1+\vartheta_2+\xi_1+\xi_2)}s_1s_2 \big| \nonumber
      \\  &=& \big|c_1s_2\pm{\rm e}^{-{\rm
      i}(\vartheta_1+\vartheta_2+\xi_1+\xi_2)}s_1c_2 \big| =
      \textstyle{1 \over \sqrt{2}} \, ,
\end{eqnarray}
and
\begin{eqnarray} \label{constraint2}
&&  \hspace*{-1.6cm} \big|c_1s_2\pm{\rm e}^{-{\rm i}(\vartheta_1 -
    \vartheta_2+\xi_1-\xi_2)}s_1c_2 \big|   \nonumber \\ &=&
    \big|c_1c_2\pm{\rm e}^{-{\rm
    i}(\vartheta_1-\vartheta_2+\xi_1-\xi_2)}s_1s_2| = \textstyle{1
    \over \sqrt{2}} \, .
\end{eqnarray}
The only solution of the constraints (\ref{constraint1}) and
(\ref{constraint2}) is 
\begin{equation} \label{con}
\cos(2\theta_1)\cos(2\theta_2)=\cos(\vartheta_1\pm\vartheta_2+\xi_1
\pm \xi_2)=0
\end{equation}
provided that neither
$\cos(2\theta_1)$ nor $\cos(2\theta_2)$ equals 1. In the special case,
where either $\cos(2\theta_1)=1$  or $\cos(2\theta_2)=1$, condition
(\ref{con})   simplifies to $\cos(2\theta_1)\cos(2\theta_2)=0$ with no
restrictions in the angles $\vartheta_1$, $\vartheta_2$, $\xi_1$ and
$\xi_2$.

One particular way to fulfill the restrictions (\ref{con}) is to set
\begin{equation} \label{angels}
\xi_2=-{\textstyle {1 \over 2}} \pi \, ,
~\xi_1=\vartheta_1=\vartheta_2=0  ~~ {\rm and} ~~\theta_1 = \theta_2 =
{\textstyle {1 \over 4}} \pi \, ,
\end{equation}
which corresponds to the choice (c.f.~\cite{moonlight})
\begin{eqnarray} \label{angelencoding}
\ket{{\sf a}_{1}} &=& {\textstyle {1 \over \sqrt{2}}} ( \ket{{\sf
    x}_0} +  \, \ket{{\sf x}_1}) \, , \nonumber \\  \ket{{\sf a}_{2}}
    &=& {\textstyle {1 \over \sqrt{2}}} ( \ket{{\sf x}_0} -  \,
    \ket{{\sf x}_1}) \, , \nonumber \\  \ket{{\sf b}_{1}} &=&
    {\textstyle {1 \over \sqrt{2}}} ( \ket{{\sf y}_0} + \ket{{\sf
    y}_1}) \, , \nonumber \\  \ket{{\sf b}_{2}} &=& {\textstyle {{\rm
    i} \over \sqrt{2}}} ( \ket{{\sf y}_0} - \ket{{\sf y}_1}) 
\end{eqnarray}
and yields
\begin{eqnarray}
\ket{\Phi_1} &= & {\textstyle {1 \over 2}}{\rm e}^{{\rm i}\pi/4} \big[
  \ket{\sf x_0y_0}-{\rm i}\ket{\sf x_0y_1}-{\rm i}\ket{\sf
  x_1y_0}+\ket{\sf x_1y_1} \big] \, , \nonumber \\  \ket{\Phi_2} &= &
  {\textstyle {1 \over 2}}{\rm e}^{-{\rm i}\pi/4} \big[ \ket{\sf
  x_0y_0}+{\rm i}\ket{\sf x_0y_1}+{\rm i}\ket{\sf x_1y_0}+\ket{\sf
  x_1y_1} \big] \, , \nonumber \\  \ket{\Phi_3} &= & {\textstyle {1
  \over 2}}{\rm e}^{{\rm i}\pi/4} \big[ \ket{\sf x_0y_0}-{\rm
  i}\ket{\sf x_0y_1}+{\rm i}\ket{\sf x_1y_0}-\ket{\sf x_1y_1} \big] \,
  , \nonumber \\  \ket{\Phi_4} &= & -{\textstyle {1 \over 2}}{\rm
  e}^{-{\rm i}\pi/4} \big[ \ket{\sf x_0y_0}+{\rm i}\ket{\sf
  x_0y_1}-{\rm i}\ket{\sf x_1y_0}-\ket{\sf x_1y_1} \big] \,
  . \nonumber \\
\end{eqnarray}
To find out which gate operation the detection of the corresponding
maximally entangled states (\ref{base}) combined with a subsequent
absorption of the photon pair results into, we write the input
state (\ref{theencoding}) as
\begin{equation} \label{bla}
\ket{\psi_{\rm enc}} = {\textstyle {1 \over 2}}  \sum_i^4
\ket{\psi_i} \otimes \ket{\Phi_i}
\end{equation} 
and determine the states $\ket{\psi_i}$ of the
stationary qubits. Using the notation
\begin{eqnarray} \label{CZ}
U_{\rm CZ} &\equiv& \ket{00}\bra{00} + \ket{01}\bra{01} +
\ket{10}\bra{10} - \ket{11}\bra{11} ~~~
\end{eqnarray}
for the controlled two-qubit phase gate (the CZ gate) and the notation
\begin{equation}
Z_i(\phi) \equiv \ket{0}_{\rm ii}\bra{0}+{\rm e}^{-{\rm
    i}\phi}\ket{1}_{\rm ii}\bra{1}
\end{equation}
for the local controlled-Z gate on photon source $i$ \cite{endnote},
we find
\begin{eqnarray}
\ket{\psi_1} &=& \exp \big({- \textstyle {1 \over 4}} {\rm i} \pi
\big) \, Z_2 \big(- {\textstyle {1 \over 2}}  \pi \big) \, Z_1 \big(-
{\textstyle {1 \over 2}}  \pi \big) \,   U_{\rm CZ} \, \ket{\psi_{\rm
in}} \, , \nonumber \\   \ket{\psi_2} &=& \exp \big({\textstyle {1
\over 4}} {\rm i} \pi \big) \, Z_2 \big({\textstyle {1 \over 2}}  \pi
\big) \,Z_1 \big( {\textstyle {1 \over 2}}  \pi \big) \, U_{\rm CZ} \,
\ket{\psi_{\rm in}} \, , \nonumber \\ \ket{\psi_3}&=& \exp \big({-
\textstyle {1 \over 4}} {\rm i} \pi \big) \, Z_2 \big(-{\textstyle {1
\over 2}}  \pi \big) \, Z_1 \big({\textstyle {1 \over 2}}  \pi \big)
\,   U_{\rm CZ} \, \ket{\psi_{\rm in}} \, , \nonumber \\
\ket{\psi_4} &=& -\exp \big({\textstyle {1 \over 4}} {\rm i} \pi
\big) \, Z_2 \big({\textstyle {1 \over 2}}  \pi \big) \, Z_1
\big(-{\textstyle {1 \over 2}}  \pi \big) \, U_{\rm CZ}   \,
\ket{\psi_{\rm in}} \, . \nonumber \\
\end{eqnarray}
From this we see that one can indeed obtain the CZ gate operation
(\ref{CZ}) up to local unitary operations upon the detection of any of
the four Bell states $\ket{\Phi_{i}}$, as it has been pointed out
already by Protsenko {\em et al.} \cite{grangier}. 

\subsection{Repeat-Until-Success quantum computing} \label{insurance}

\noindent
When implementing  distributed quantum computing with photons as
flying qubits,  the problem arises that it is impossible to perform a
complete deterministic Bell  measurement on the photons using only
linear optics elements. As it has been shown 
\cite{Lutkenhaus}, in the best case, one can distinguish two of the
four Bell states. Since the construction of efficient non-linear optical
elements remains experimentally challenging, the above described phase 
gate could therefore be operated at most with success rate ${1 \over 2}$. 

What must be done to solve this problem is to choose the photon pair measurement 
basis $\{ |\Phi_i \rangle \}$ such that two of the basis states are
maximally entangled while the other two basis states are product
states. Most importantly, all basis states must be mutually unbiased
with respect to the computational basis and information will not be
destroyed at any   stage of the computation. In the following we
choose $\ket{\Phi_3}$ and  $\ket{\Phi_4}$ as in Eq.~(\ref{base}) and
$\ket{\Phi_1}$ and $\ket{\Phi_2}$ as product states such that
\begin{eqnarray}\label{twoproduct}
\ket{\Phi_1} &=& \ket{{\sf a_1b_1}} \, , ~~  \ket{\Phi_2}=\ket{{\sf
    a_2b_2}} \, , \nonumber \\ 
\ket{\Phi_3} &\equiv & {\textstyle {1 \over
      \sqrt{2}}} \big[ \ket{{\sf a}_1 {\sf b}_2}+\ket{{\sf a}_2 {\sf
      b}_1} \big] \, , \nonumber \\ \ket{\Phi_4} &\equiv & {\textstyle
      {1 \over \sqrt{2}}} \big[ \ket{{\sf a}_1 {\sf b}_2}-\ket{{\sf
      a}_2 {\sf b}_1} \big] \, .
\end{eqnarray}
The aim of this is (see Section \ref{sectionii}) that in the event of
the ``failure'' of the gate implementation (i.e.~in case of the detection of
$\ket{\Phi_1}$ or $\ket{\Phi_2}$) the system remains, up to a local
phase gate, in the original qubit state. This means that the initial 
state (\ref{original}) can be restored and the described protocol can
be repeated, thereby eventually resulting in the performance of the
universal controlled phase gate (\ref{CZ}). The probability for the
realization of the gate operation within one step equals ${1 \over 2}$ and 
the final completion of a quantum phase gate therefore requires on
average {\em two} repetitions of the above described photon pair
generation and detection process. 

Let us now determine the conditions under which the states $\{ |\Phi_i
\rangle \}$ are of the form (\ref{unbiased}). Proceeding as above,
we find that the angles $\vartheta_i$, $\xi_i$ and $\theta_i$ in
Eq.~(\ref{ab}) should fulfill, for example, Eq.~(\ref{angels}). In
analogy to Eqs.~(\ref{constraint1}) and (\ref{constraint2}), we find
that $\ket{\Phi_1}$ and $\ket{\Phi_2}$ are mutually unbiased, if
\begin{eqnarray}
\big|c_1c_2 \big| = \big| c_1s_2 \big| = \big|s_1c_2 \big| =
\big|s_1s_2 \big| = \textstyle{1 \over 2} \, ,
\end{eqnarray}
which also holds for the parameter choice in Eq.~(\ref{angels}). 
Using Eq.~(\ref{angelencoding}), one can
easily verify that with the above choice the basis (\ref{twoproduct})
becomes 
\begin{eqnarray} \label{xxx}
\ket{\Phi_1}&=& {\textstyle {1 \over 2}} \big[ \ket{\sf
    x_0y_0}+\ket{\sf x_0y_1}+\ket{\sf x_1y_0}+\ket{\sf x_1y_1} \big]
    \, , \nonumber \\  \ket{\Phi_2}&=& {\textstyle {{\rm i} \over 2}}
    \big[ \ket{\sf x_0y_0}-\ket{\sf x_0y_1}-\ket{\sf x_1y_0}+\ket{\sf
    x_1y_1} \big] \, , \nonumber \\
\ket{\Phi_3} &= & {\textstyle {1
  \over 2}}{\rm e}^{{\rm i}\pi/4} \big[ \ket{\sf x_0y_0}-{\rm
  i}\ket{\sf x_0y_1}+{\rm i}\ket{\sf x_1y_0}-\ket{\sf x_1y_1} \big] \,
  , \nonumber \\  \ket{\Phi_4} &= & -{\textstyle {1 \over 2}}{\rm
  e}^{-{\rm i}\pi/4} \big[ \ket{\sf x_0y_0}+{\rm i}\ket{\sf
  x_0y_1}-{\rm i}\ket{\sf x_1y_0}-\ket{\sf x_1y_1} \big] \, . \nonumber \\
\end{eqnarray} 
Choosing the states $|{\sf a}_i \rangle$ and $|{\sf
b}_i  \rangle$ as in Eq.~(\ref{angelencoding}) allows to implement the
gate operation (\ref{CZ}) eventually deterministically.

Finally, we determine the gate operations corresponding to the
detection of a certain measurement outcome $|\Phi_i \rangle$. To do
this, we decompose the input state (\ref{theencoding}) again into a state
of the form (\ref{bla}). Proceeding as in the previous subsection, we
find
\begin{eqnarray} \label{PossibleOutcomesOfCZ}
\ket{\psi_1} &=&  \ket{\psi_{\rm in}} \, , \nonumber \\   \ket{\psi_2}
&=& -{\rm i} \, Z_2 \big( \pi \big) \, Z_2 \big( \pi \big) \,
\ket{\psi_{\rm in}} \, , \nonumber \\  \ket{\psi_3}&=& \exp \big({-
\textstyle {1 \over 4}} {\rm i} \pi \big) \, Z_2 \big(-{\textstyle {1
\over 2}}  \pi \big) \, Z_1 \big({\textstyle {1 \over 2}}  \pi \big)
\,   U_{\rm CZ} \, \ket{\psi_{\rm in}} \, , \nonumber \\
\ket{\psi_4} &=& -\exp \big({\textstyle {1 \over 4}} {\rm i} \pi
\big) \, Z_2 \big({\textstyle {1 \over 2}}  \pi \big) \, Z_1
\big(-{\textstyle {1 \over 2}}  \pi \big) \, U_{\rm CZ}   \,
\ket{\psi_{\rm in}} \, . \nonumber \\
\end{eqnarray}
Again one obtains the CZ gate operation
(\ref{CZ}) up to local unitary operations upon the detection of either
$\ket{\Phi_{3}}$ or $\ket{\Phi_{4}}$. In the event of the detection of the
product states $\ket{\Phi_1}$ or $\ket{\Phi_2}$, the initial state can
be restored with the help of one-qubit phase gates, which then allows us
to repeat the operation until success.

It should be emphasized that there are other possible encodings that
yield a universal two-qubit phase gate upon the detection of a
Bell-state, but where the original state is destroyed upon the
detection of a product state (see e.g.~\cite{Zou}). This happens when
the product states are not mutually unbiased and their detection
erases the qubit states in the respective photon sources. To achieve the effect of
an {\em insurance} against failure, the encoding (\ref{enc}) should be
chosen as described in this Section.

\section{Possible experimental realizations} \label{rea}

\noindent
Possible experimental realizations of the above described eventually
deterministic  quantum phase gate consist of two basic steps. Firstly,
the information of the stationary qubits involved in the operation has
to be redundantly encoded in the states of two newly generated ancilla
photons. Afterwards, a measurement is performed on the photon pair
resulting with probability ${1 \over2}$ in the desired gate
operation. Depending on the type of the photon source, one can choose
different types of encoding. There are also different possibilities
how to perform the photon pair measurement.  Examples are given below.

\subsection{Redundant encoding} 

\begin{figure}
\begin{minipage}{\columnwidth}
\begin{center}
\resizebox{\columnwidth}{!}{\rotatebox{0}{\includegraphics{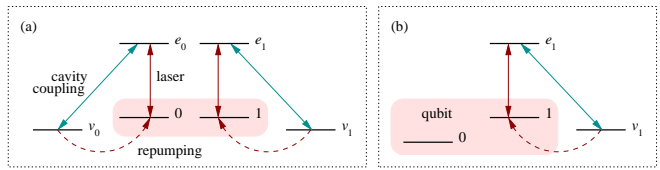}}}
\end{center}
\vspace*{-0.5cm}
\caption{Schematic view of a single photon (a) polarization encoder, 
Ê(b) time-bin encoder and level configuration of the sourceÊ 
Êcontaining the qubit.} \label{photongun}
\end{minipage}
\end{figure}

\noindent 
In order to obtain robust qubits, the states $|0 \rangle$ and $|1
\rangle$ should be two different longliving ground states of the
single photon source. Each photon source carries one qubit. Depending
on its level structure (see Figure \ref{photongun}), it might be
advantageous to realise the encoding step (\ref{theencoding}) either
by generating photons with different polarisations (polarisation
encoding) or photons that agree in all degrees of freedom apart from
their creation time (time bin encoding). Note that different encodings
can easily be transformed into each other using linear optics elements
like a polarising beam splitters and delaying photons in time. 

{\em Polarization encoding.}   Suppose, the photon source contains an
atomic double $\Lambda$ level configuration as shown in Figure
\ref{photongun}(a) (see also Ref.~\cite{Gheri}).  A single photon can
then be created by simultaneously applying a laser pulse with
increasing Rabi frequency to the $0$-$e_0$ transition and the
$1$-$e_1$ transition of the atomic system. Thereby, the atom goes to
the ground state $|v_0 \rangle$ and $|v_1 \rangle$, respectively,
depending on whether its initial state  equalled $|0 \rangle$ or $|1
\rangle$ due to the coupling of the $e_0$-$v_0$ transition and the
$e_1$-$v_1$ transition to the cavity mode. It has been shown in the
past that this technique \cite{Law} is very well suited to place
exactly one excitation into the field of an optical resonator, from
where it can leak out \cite{Kuhn2}.

If the two transitions, $e_0$-$v_0$ and $e_1$-$v_1$, couple to the two
different polarisation modes ${\sf h}$ and ${\sf v}$, in the cavity
field, the photon generation results effectively, for example, in the
operation 
\begin{equation} 
\ket{0}_i \rightarrow \ket{0 ,{\sf h}}_i \, , ~~ \ket{1}_i \rightarrow
\ket{1 ,{\sf v}}_i  
\end{equation}
once atom $i$ has been repumped into its initial state $|0 \rangle_i$
and $|1 \rangle_i$, respectively. Finally we remark that the encoding
does not affect the coefficients $\alpha$, $\beta$, $\gamma$ and
$\delta$ of the initial state (\ref{original}). As long as no
measurement is performed on the system, all coherences are preserved. 

{\em Time-bin encoding.} 
Alternatively, if the photon sources possess a level structure like
the one shown in Figure \ref{photongun}(b), one can redundantly encode
the information contained in the qubits into time bin encoded photons, 
\begin{equation} \label{higgs}
\ket{0}_i \rightarrow \ket{0 ,{\sf E}}_i \, , ~~ \ket{1}_i \rightarrow
\ket{1 ,{\sf L}}_i \, .  
\end{equation}
This encoding is simpler and may therefore find realizations not only in
atoms but also in quantum dots and nitrogen vacancy color centers. 
In Eq.~(\ref{higgs}), $|{\sf E} \rangle$ and $|{\sf L} \rangle$ denote
a single photon generated at an {\em early} and a {\em later} time,
respectively. The above operation can be achieved by first coupling a
laser field with increasing Rabi frequency to the $1-e_1$ transition,
while the cavity mode couples to the $e_1-v_1$ transition. Once the
excitation has been placed into the cavity mode and leaked out through
the outcoupling mirror, the atom can be repumped into $|0
\rangle$. Afterwards, one should swap the states $\ket{0}$ and
$\ket{1}$ and repeat the process. This results in the generation of a
late photon, if the system was initially prepared in $|1 \rangle$. To
complete the encoding, the states $\ket{0}$ and $\ket{1}$ have to be
swapped again. 

\subsection{Photon pair measurement}\label{rea2}

\noindent
We now give two examples how to perform a photon pair measurement of
the mutually unbiased basis (\ref{xxx}). The first method is suitable
for polarization  encoded photons, the second one for dual-rail
encoded photons. If the qubits have initially been time bin-encoded,
their encoding should be transformed first using standard linear
optics techniques.

\begin{figure}\label{moon2}
\begin{minipage}{\columnwidth}
\begin{center}
\resizebox{\columnwidth}{!}{\rotatebox{0}{\includegraphics{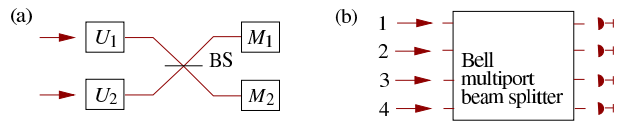}}}
\end{center}
\vspace*{-0.5cm}
\caption{Linear optics networks for the realization of a measurement
  of the basis states (\ref{base}) after encoding the photonic qubits
  in the polarization degrees of two photons (a) or into four
  different spatial photon modes (b) involving either a beam splitter
  or a $4 \times 4$ Bell multiport beam splitter.} \label{moon3}
\end{minipage}
\end{figure}

{\em Polarization encoding.} 
It is well known that sending two
polarization encoded photons through the different input ports of a
50:50 beam splitter together with polarization sensitive measurements
in the $\ket{\sf h}/\ket{\sf v}$-basis in the output ports would
result in a measurement of the states ${\textstyle {1 \over {\sqrt
2}}}(\ket{\sf hv}\pm\ket{\sf vh})$, $\ket{\sf hh}$ and $\ket{\sf vv
}$. To measure the states (\ref{twoproduct}), we therefore propose to
proceed as shown in Figure \ref{moon3}(a) \cite{moonlight} and to
perform the mapping 
\begin{eqnarray}
U_1 &=& |{\sf h} \rangle \langle {\sf a_1}| + |{\sf v} \rangle \langle
{\sf a_2}| \, , \nonumber \\ 
U_2 &=& |{\sf h} \rangle \langle {\sf b_1}| + |{\sf v} \rangle \langle
{\sf b_2}| 
\end{eqnarray} 
on the photon coming from source $i$. Using Eq.~(\ref{angelencoding}),
we see that this corresponds to the single qubit rotations
\begin{eqnarray}
U_1 &=&  {\textstyle {1 \over \sqrt{2}}} \, \big[ \, |{\sf h} \rangle
  \big( \langle {\sf h}| + \langle {\sf v}| \big) +  |{\sf v} \rangle
  \big( \langle {\sf h}| - \langle {\sf v}| \big) \, \big] \, ,
  \nonumber \\ U_2 &=&  {\textstyle {1 \over \sqrt{2}}} \, \big[ \,
  |{\sf h} \rangle \big( \langle {\sf h}| + \langle {\sf v}| \big) -
  {\rm i} \, |{\sf   v} \rangle \big( \langle {\sf h}| -  \langle {\sf
  v}| \big) \, \big] \, .
\end{eqnarray}
After leaving the beam splitter, the photons should be detected in the
$\ket{\sf h}/\ket{\sf v}$-basis. A detection of two ${\sf h}$  and two ${\sf
v}$ polarized photons indicates a measurement of $|\Phi_1 \rangle$
and $|\Phi_2 \rangle$, respectively. Finding two photons of different
polarization in the same or in different detectors corresponds to a
detection of $|\Phi_3 \rangle$ or $|\Phi_4 \rangle$.

{\em Dual-rail encoding.}  
Alternatively, one can redirect the generated photons to the different
input ports of a $4 \times 4$ Bell multiport beam splitter as shown in
Figure \ref{moon3}(b). If $a_n^\dagger$ and $b_n^\dagger$ denotes the
creation operator for a photon in input and output port $n$,
respectively,  the effect of the multiport can be summarized as
\cite{multi} 
\begin{equation} \label{scatter}
a_n^\dagger \to \sum_m U_{mn} b_m^\dagger 
\end{equation}
with
\begin{equation}
U_{mn} = {\textstyle {1 \over 2}} \, {\rm exp} \big( {\rm i} \pi
(n-1)(m-1) \big) \, . 
\end{equation}
A Bell multiport redirects each incoming photon with equal probability
to any of the possible output ports, thereby erasing
the which-way information of the incoming photons. One way to measure
in the mutually unbiased basis (\ref{xxx}) is to direct the $\ket{\sf
x_0}$ photon from source 1 to input port 1, the $\ket{\sf x_1}$ photon
from source 1 to input port 3 and to direct the $\ket{\sf y_0}$ photon
and the  $\ket{\sf y_1}$ photon from source 2 to input port 2 and 4,
respectively. If $\ket{\rm vac}$ denotes the state with no photons in
the setup, this results in the conversion  
\begin{eqnarray}
&& |{\sf x_0y_0} \rangle \to a_1^\dagger a_2^\dagger \, \ket{{\rm vac}} \, , ~~
|{\sf x_0y_1} \rangle \to a_1^\dagger a_4^\dagger \, \ket{{\rm vac}}
\, , \nonumber \\ 
&& |{\sf x_1y_0} \rangle \to a_2^\dagger a_3^\dagger \, \ket{{\rm vac}} \, , ~~
|{\sf x_1y_1} \rangle \to a_3^\dagger a_4^\dagger \, \ket{{\rm vac}} \, . 
\end{eqnarray}
This conversion should be realized such that the photons enter
the multiport at the same time. Using Eq.~(\ref{scatter}) one can show
that the network transfers the basis states (\ref{xxx}) according to
\begin{eqnarray} \label{multiportresult}
|\Phi_1 \rangle & \to &  {\textstyle {1 \over 2}} \, \big(
b_1^{\dagger \, 2}  - b_3^{\dagger \, 2} \big) \,  \ket{\rm vac} \, ,
\nonumber \\   |\Phi_2 \rangle & \to & - {\textstyle {1 \over 2}} \,
\big( b_2^{\dagger \, 2}  - b_4^{\dagger \, 2} \big) \, \ket{\rm vac}
\, , \nonumber \\  |\Phi_3 \rangle & \to & {\textstyle {1 \over {\sqrt
      2}}} \, \big( b_1^\dagger b_4^\dagger - b_2^\dagger
b_3^\dagger \big) \, \ket{\rm vac} \, , \nonumber  \\    |\Phi_4
\rangle & \to & - {\textstyle {1 \over {\sqrt 2}}} \, \big(
b_1^\dagger b_2^\dagger - b_3^\dagger   b_4^\dagger  \big) \, \ket{\rm
  vac} \, .
\end{eqnarray}
Finally, detectors measure the presence of photons in each of the
possible output ports. The detection of two photons in the same output
port, namely in 1 or 3 and in 2 or 4, corresponds to a measurement of
the state $|\Phi_1 \rangle$ and $|\Phi_2 \rangle$, respectively. The
detection of a photon in ports 1 and 4 or in 2 and 3 indicates a
measurement of the state $\ket{\Phi_3}$, while a photon in the ports 1
and 2 or in 3 and 4 indicates the state $\ket{\Phi_4}$.

Any unknown fixed (or slowly varying with respect to the coherence length of
the photon pulse) phase factor introduced along the photon paths contributes
at most to a global phase factor to the input state (\ref{theencoding}), which
is  also a feature of the schemes outlined in
Refs.~\cite{last,Simon,lim,moonlight}. The implementation of
Repeat-Until-Success quantum computing therefore does not require
interferometric stability. It requires only overlapping of the photons within
their coherence length within the linear optics setup.

\section{Scalable quantum computation in the presence of
inefficient photon generation and detection}\label{cluster}

\noindent
In this section, we discuss the possibility of implementing scalable
quantum computation using the Repeat-Until-Success quantum gate
described in the previous sections. The implementation of this gate
requires the generation of single photons on demand and linear optical
elements together with absorbing quantum measurements. In the limit
of perfect photon emission, collection, and detection efficiency,
two-qubit CZ gates can be performed deterministically, as described
above. In real systems however, photon emission, collection, and
detection is not perfect \cite{kok01}. In existing experiments, all
of these processes have significant inefficiencies, which means that
there is a finite probability that two photons will not be observed in
the photon measurement. The failure to observe two photons in an
attempted CZ operation means that the static qubits are left in an
unknown state, which constitutes a correlated two qubit error. If such
losses are sufficiently small (e.g. less than $\sim 10^{-2}$), the
resulting gate failures can be dealt with using existing fault
tolerance techniques \cite{steane2003,knill2005}. Recently, much
higher fault tolerance levels of up to 50\% were found in linear
optical quantum computing \cite{varnava,gilchrist}.

More concretely, the highest reported photon detection efficiency for
single photon detection with photon number resolution is about $88\%$
\cite{Takeuchi99,Rosenberg}. A recent experiment by McKeever {\em
et al.} \cite{Mckeever} involving an atom-cavity system for
the generation of single photons yields a photon generation efficiency of
nearly 70\%, limited only by passive cavity loss. The lifetime
of the atom in the cavity was $0.14\,$s, allowing for as many as $1.4 \cdot 10^4$ photon
generation events. Moreover, Legero {\em et al}.~\cite{Legero} demonstrated perfect time-resolved interference with two photons of different frequencies. Time-resolved detection acts as a temporal
filter to erase the which-way information which is important to any scheme
involving photon interference. This suggests that strictly
identical single photon sources are not required for attaining high fidelities
in the state preparation. The cost of this high fidelity is a lower
probability of success.  

Fortunately, scalable quantum computing is possible, even in the presence of
large errors, as long as no errors imply a very high fidelity and the occurrence of an error is \emph{heralded}:
if fewer than two photons are detected, we know that the attempted CZ operation has failed. Only when the detectors have a substantial amount of dark
counts, we cannot rely on this error detection mechanism. However,
comercially available silicon avalanche photodetectors are available with a
detection efficiency of 65\%, and a dark count rate of $\Gamma_{dc} \le
25\,$s$^{-1}$ \cite{PED}. Assuming, a photon regeneration  rate of
$105\,$s$^{-1}$ gives a clock time of $10^{-5}$ s. The total dark count
probability is then $p_{dc} \approx 10^{-4}$ per clock cycle, which is small
enough to be dealt with using existing  error correction techniques.
Moreover, if one could experiment with detectors like the one reported by
Rosenberg {\em et al}.~\cite{Rosenberg}, dark count rate effects would be 
negligible.

In the case of
an error, the state of the static qubits can be determined by
subsequently performing measurements on the sources, which allows the
sources to be re-prepared in a known state. In earlier work, we have
shown that scalable quantum computation can be performed in the
presence of significant heralded error rates, by first using a
non-deterministic entangling operation to create \emph{cluster states}
of many qubits \cite{Sean}, and subsequently implementing scalable
quantum computation via the `one-way quantum computer'
\cite{raussendorf}. Given cluster states of many qubits, the one-way
quantum computer can be implemented by single qubit measurements
alone. This technique permits fully scalable quantum computation,
albeit with a fixed overhead per two qubit gate in the algorithm,
which we calculate below. We briefly review how one way quantum
computing can proceed within our scheme, and then provide an estimate
of the overhead costs involved.

\subsection{One way quantum computation}

\begin{figure}[t]
   \begin{center}
        \epsfig{file=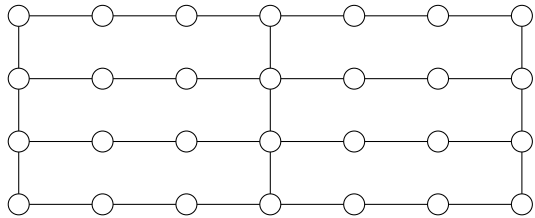, width=7cm}
   \end{center}
    \caption{A rectangular lattice cluster state. Each circle
    represents a physical qubit, and each line represents a `bond'
    between qubits. These states are sufficient for simulating
    arbitrary logic networks \cite{nielsen04}, with each horizontal row
    representing a single logical qubit, and each vertical connection
    representing a two-qubit gate. Note we also permit bonds 
    between non-adjacent rows and columns (not shown), which can
    simulate non-nearest neighbor two-qubit operations.}
    \label{fig:cluster1}
\end{figure}

\noindent
One way quantum computation \cite{raussendorf} proceeds by first
creating a graph state of many qubits, and subsequently performing
single qubit measurements on the graph state \cite{rbb,hein,weinstein}. Graph
states may be  represented as a graph comprising set of qubit `nodes'
connected by `edges' which may be understood as `bonds' between the
qubits.  The quantum state corresponding to such a graph may be
defined (and also implemented) by the following procedure: (i) prepare
each qubit in the state $|+\rangle = (|0\rangle +
|1\rangle)/\sqrt{2}$, and then (ii) for each bond in the corresponding
graph, apply a deterministic CZ operation (see Eq.  \ref{CZ}) between
the relevant qubits. In this work, we will restrict our attention to
the rectangular lattice graph states of the form shown in Figure
\ref{fig:cluster1} (hereafter referred to simply as \emph{cluster
states}), which are sufficient for simulating arbitrary logic
networks, and hence universal quantum computations
\cite{nielsen04}. It is worth noting however that straightforward
generalizations of the procedure described below allow us to scalably
generate \emph{arbitrary} graph states. This may be useful in that it
might result in reduced costs for implementing certain
algorithms. 

In these clusters, each horizontal row of physical qubits
represents a single logical qubit in the logic network being
simulated. Two qubit operations are implemented by the vertical
bonds acting between rows. We also permit bonds between
non-adjacent rows, which permits highly non-local two qubit gates
to be implemented. Note also that the location of the qubits
within the cluster is notional, and need not correspond to the
physical location of the static qubit (the mapping between the
notional qubit positions within the cluster, and the actual
physical location of the qubits can be stored in a classical
computer). After making the state, quantum computation proceeds by
performing a sequence of single qubit measurements on the static
qubits, with each measurement performed in a particular basis so
as to implement a given sequence of one- and two-qubit gates
\cite{raussendorf,nielsen04}. At each time step, a whole column of
physical qubits in the cluster is measured. The measurements are
performed in order, starting with the column at left side of the
cluster, and proceeding rightwards across the cluster. In general,
the basis of the measurements made at a given time step will
depend on the outcomes of earlier measurements. Once a physical
qubit has been measured, that qubit is disentangled from the
cluster state and so may be re-initialized in a particular state
and subsequently used later in the computation.

We assume that single-shot single qubit measurements and single
qubit unitary operations on the static qubits can be implemented
using standard techniques. Implementing one-way quantum
computation in our scheme therefore reduces to the problem of
scalably generating cluster states using the heralded,
non-deterministic CZ operation. We outline the general procedure
here, and give a more detailed description in the subsequent
section.

In our scheme, cluster states can be generated by attempting to
bond qubits using the non-deterministic CZ operation. This
operation has three possible outcomes: `success', `insurance', or
`failure'. In the case of observing two photons, one of the gates
of Eq.(\ref{PossibleOutcomesOfCZ}) is implemented, and subsequent
application of appropriate single qubit unitaries implements
either the CZ operation (denoting a `success'), or the identity
operation (denoting `insurance'). In the case of insurance, the CZ
operation can simply be reattempted. Observing fewer than two
photons denotes a failure. In this case, the static qubits are
left in an unknown state. However, this damage can be repaired as
follows. Firstly, each of the two qubits involved in the failed
gate can be measured in the computational basis to determine the
nature of the error. If either qubit was already part of a cluster
state, the bonds to its neighbors within the cluster are also
destroyed. However, the remainder of the cluster state can be
recovered by applying appropriate single qubit unitary operations
to these neighboring qubits, conditional on the outcome of the
measurement on the qubit involved in the failed CZ gate.
Therefore, the cluster state can grow, shrink, or remain the same
size, depending whether the CZ operation was successful, failed,
or failed with insurance. The key to scalably generating cluster
states is to attempt CZ operations between qubits in a sequence
order such that the cluster state grows on average. We give such a
sequence in Sec. \ref{Overhead costs}.

\begin{figure}[t]
   \begin{center}
   \begin{psfrags}
      \psfrag{M}{M}
      \psfrag{L}{$L_a$}
      \psfrag{K}{$L_a$}
      \psfrag{buffer}{buffer}
        \epsfig{file=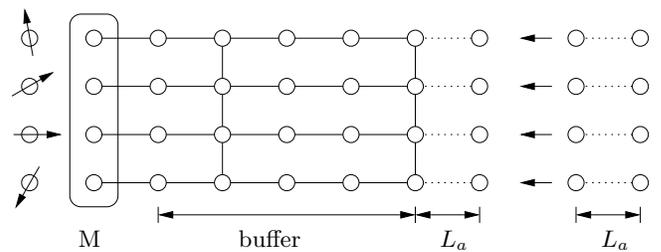, width=8.5cm}
   \end{psfrags}
   \end{center}
    \caption{Dynamically growing clusters during a computation. The
      cluster contains three regions: the active region (M), at the
      left of the cluster, in which the logic gate networks are being
      simulated via single qubit measurements; the buffer region; and
      the connection region, where new cluster fragments are added to
      the right edge of the main cluster. The connecting cluster chains
      have a buffer length $L_a$ to accommodate the probabilistic
      entangling operation.}
    \label{fig:cluster2}
\end{figure}

We conclude this section by noting that it is not necessary
to build the whole cluster required for simulating a
particular algorithm before commencing the single qubit
measurement part of the computation. It is possible to
build a partial cluster, and then to simultaneously perform
single qubit measurements on one part of the cluster, while
adding new qubits to another region in the cluster. In this
approach to one-way quantum computing, one can think of the
cluster as being split into three regions, as shown in Figure
\ref{fig:cluster2}. The \emph{active region}, to the left
of the cluster, contains the part of the cluster where the
logic gate networks are being simulated via single qubit
measurements. At the right of the cluster, the
\emph{connection region} comprises of several horizontal
dangling linear chains which extend from the right edge of
the main cluster, each corresponding to a logical qubit. In
this region, nondeterministic CZ operations are applied in
order to add further cluster sections to the main section.
These additional sections are manufactured separately, as
described in Sec. \ref{Overhead costs}. Between the active
region and the connection region, the \emph{buffer region}
comprises a quiescent region which suffices to protect the
active region in the event of a long sequence of failed CZ
operations; this would lead to the right edge of the
cluster running back into the active region, damaging the
logical computation. The depth of the buffer region should
be chosen such that the probability of erasing a logical
qubit is sufficiently small that it can be handled with
existing fault tolerance techniques
\cite{steane2003,knill2005}.

There are several advantages to this approach. Firstly,
fewer physical qubits are needed, because qubits that have
already been measured at the left edge of the cluster can
be recycled and added to the right hand side of the
cluster. Secondly, preparing the whole cluster initially
means that some of the qubits will spend a lot of time in
an `idle' state before they are involved in the
computation; any errors accumulated in these idle qubits
due to decoherence will degrade the fidelity of the
computation \cite{raussendorf}. This is crucial if fault
tolerant quantum computation is to be implemented within
the cluster model, as such schemes require a source of
fresh ancilla qubits throughout the algorithm. Thirdly, the
overhead costs for this approach can be reduced, because it
is not necessary to prepare the \emph{whole} cluster with a
total success probability close to one; the probability for
erasing a given logical qubit need only be made smaller
than the error threshold required for fault tolerance.

\subsection{Overhead costs} \label{Overhead costs}

\noindent
A number of authors have considered efficient cluster state generation
using non-deterministic, but heralded, Entangling Operations (EOs)
\cite{yoran03,nielsen04,browne04,Benjamin,Sean,DuanRaussendorf,benjamin04,chen}.
References \cite{yoran03,nielsen04,browne04} calculated explicit costs
for making cluster states of optical qubits in the ideal case
(i.e. neglecting photon loss). Subsequently, Barrett and Kok
\cite{Sean} showed that, in the case of hybrid matter-optical systems
(such as those considered in this work), arbitrarily small EO success
probabilities could be tolerated. They provided a `divide and conquer'
algorithm for building linear clusters, which has moderate costs even
for small success probability. An efficient algorithm for building two
dimensional clusters, capable of simulating arbitrary logic networks
was also given in \cite{Sean}. More recently, in
Ref. \cite{DuanRaussendorf}, a similar algorithm for building linear
clusters was proposed, which made more use of recycling, and hence has
a lower overhead cost.  Ref. \cite{DuanRaussendorf} also gives an
alternative algorithm for making 2-dimensional clusters, and
explicitly calculates the associated overhead costs. In
Ref. \cite{benjamin04}, some elegant cost reducing improvements to the
scheme proposed in Ref.  \cite{Sean} were suggested, utilizing the
redundantly encoded qubits inherent in the original scheme.

In this work, we will combine elements of the approaches taken in
Refs. \cite{browne04,Sean,DuanRaussendorf} to provide a simple upper
bound for the scaling costs for building cluster states using our
scheme. This estimate is based on an explicit procedure, and we do not
claim that it is optimal; an improved algorithm may yield
substantially reduced costs. Nevertheless, the procedure given here
allows a straightforward calculation of the overhead costs.  Despite
its apparent similarity to Refs.~\cite{Sean,DuanRaussendorf}, there is
a crucial difference: in the scheme under consideration in this paper,
there is the possibility of obtaining the `insurance' outcome. In
general, this leads to a reduction in costs relative to schemes in
which there is no insurance outcome.

In the presence of imperfect photon emission, detection, and
collection, the performance of the CZ operation can be
characterized by three probabilities:
\begin{itemize}
\item the probability of successfully implementing the CZ
operation on the input qubits (up to local operations),
$p_s$;
\item  the probability of obtaining the `insurance' outcome
in which known local operations are applied to the qubits,
$p_i$;
\item the probability of failure due to failing to emit, collect,
or detect one or more photons during the remote gate operation,
$p_f$.
\end{itemize}
These probabilities are determined by the physics of the sources
and detectors.

Calculating the total cost of growing cluster states can be
simplified by noting that, in the case of obtaining the
`insurance' outcome, after applying the necessary single qubit
corrections, one simply attempts the gate operation again. This
process is repeated until a definite outcome (success or failure)
is obtained. Thus, we can define \emph{total} success and failure
probabilities, $P_s$ and $P_f$, of the corresponding definite
outcomes after an (arbitrarily long) sequence of `insurance'
outcomes. These probabilities are given by $P_s =
\sum_{j=0}^{\infty} p_i^j p_s = p_s/(1-p_i)$ and $P_f =
\sum_{j=0}^{\infty} p_i^j p_f = p_f/(1-p_i)$. The average number
of attempted CZ operations required before we obtain a definite
outcome is $N_{\mathrm{av}} = 1/(1-p_i)$.

The overhead cost for making cluster
states is then found using similar calculations to those presented
in Refs. \cite{Sean,DuanRaussendorf}. We first calculate the cost
(i.e. the number of attempted CZ operations per qubit in the final
cluster) of generating linear clusters. If a CZ gate is repeatedly
applied between the end qubits of two linear chains, each of
length $L_k$, either the gate is (ultimately) successful, in which
case the total length of the new cluster is $2 L_k$, or the gate
(ultimately) fails, in which case, the length of the original
clusters shrinks by one qubit each. Repeatedly applying this
procedure until a successful outcome is obtained (or until both
original clusters are destroyed) \cite{DuanRaussendorf} gives the
expected length $L_{k+1} = \sum_{i=0}^{L_k} 2 (L_k - i) P_s P_f^i
\approx 2 L_k - 2p_f/p_s$. Denoting the average number of attempts
to create a chain of length $L_k$ by $N_k$, we also have $N_{k+1}
= 2 N_k + 1/p_s$. Solving these recursion relations gives a total
cost
\begin{equation}
N(L) = \frac{\left(N_0+ \frac{1}{p_s}\right) \left(L -
\frac{2p_f}{p_s}\right) }{ \left(L_0 - \frac{2p_f}{p_s} \right)} -
\frac{1}{p_s}, \label{TotalCost}
\end{equation}
where $N_0$ denotes the cost of growing a short cluster of length
$L_0$. Note that for the average cluster length to grow on each
round of the protocol, we require $L_1 > L_0$, which implies that
the length of the short chains should satisfy $L_0 > 2p_f/p_s$.

\begin{figure}[t]
   \begin{center}
   \begin{psfrags}
      \psfrag{a}{a)}
      \psfrag{b}{b)}
      \psfrag{c}{c)}
      \psfrag{d}{d)}
      \psfrag{X}{$\sigma_x$}
      \psfrag{H}{\footnotesize $H$}
         \epsfig{file=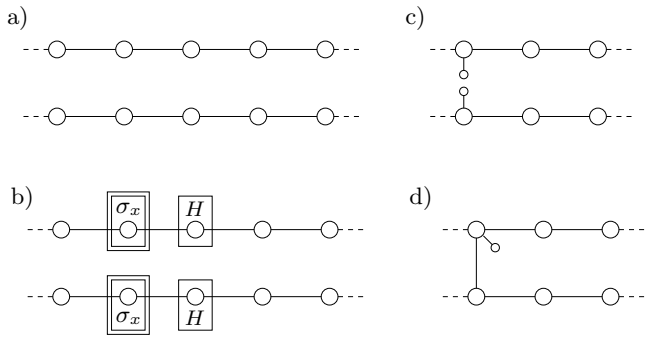, width=8.5cm}
   \end{psfrags}
   \end{center}
    \caption{Creating vertical bonds. a) We start out with two
    sufficiently long cluster chains, and we wish to create a vertical
    bond between the two qubits on the left. b) We apply a $\sigma_x$
    measurement to the two adjacent qubits and a Hadamard operation
    $H$ on the next. c) This will result in a
    redundant encoding of the qubits we wish to bond together. d)
    Applying the entangling operation to the dangling ``cherries''
    will create the vertical bond. Note that we also removed the
    qubits in the vertical bond by applying another $\sigma_x$
    measurement, resulting in another redundant encoding. If this
    procedure fails, we are left with a shorter chain, and we can try
    to create a vertical bond again.}
    \label{fig:cluster3}
\end{figure}

Chains of fixed length $L_0$ can be grown independently using the
probabilistic CZ operation, by joining sub-chains together.
Growing these short chains adds a constant overhead cost to the
cluster generation process. We use a `divide and conquer' approach
to making these short chains \cite{Sean,DuanRaussendorf}, in
which, on each round of the protocol, we attempt to join equal
length pairs of linear clusters using the probabilistic CZ
operation. If we obtain the `insurance' outcome on any such
attempt, we try the operation again, whereas if we fail, we assume
(for ease of calculation) that the short chains are discarded. On
the $k$th round of this protocol, the length of the chains is $l_k
= 2^k$, and the number of attempted CZ operations is given by the
recursion relation $n_k = 2 n_{k-1}/P_s + N_{\mathrm{av}}/P_s$.
Solving these relations gives
\begin{equation}
N_0(L_0) = N_{\mathrm{av}}\sum_{i=1}^{\log_2 L_0}
\frac{2^{i-1}}{P_s^i} \,. \label{OffLineCost}
\end{equation}
Combining Eq. (\ref{OffLineCost}) and Eq.
(\ref{TotalCost}), one can calculate the total cost of
growing linear clusters for given values of $p_f$, $p_s$
and $p_i$. For instance, taking $p_f = 0.6$, $p_i = p_s =
0.2$, we require $L_0 > 6$. Taking $L_0 = 2^3 = 8$, the
total cost for making a linear cluster of length $L$ is
found to be $N(L) = 185 L - 1115$ attempted CZ operations.
A moderate increase in success probability can dramatically
decrease the cost: taking $p_f = 0.4$, $p_i = p_s = 0.3$,
we require $L_0 > 2.67$, and taking $L_0=2^2=4$, we find
the total cost to be $N(L) = 16\frac{2}{3} L -
47\frac{7}{9}$. Note that the negative constant term in
these expressions is an artefact of joining small numbers
of chains together to make an isolated chain of length $L$.
This is an `edge effect' which should be neglected when
considering the asymptotic cost of making long chains.

\begin{table}[t]
 \caption{The average number of entangling operations per vertical
 bond, given by $N_{\rm bond} = 2 N(M) + (1-p_i)/p_s$. Here $p_s$,
 $p_f$ and $p_i$ are the success, failure and insurance probabilities,
 respectively. $L_0=2^n$ is the length of the chain that is needed to
 obey the growth requirement, and $N_0$ is the number of EOs needed to
 achieve this length. $M$ is the average cluster chain consumed by the
 forging of a vertical bond, and $N(M)$ is the number of EOs needed to
 achieve this length.}
 \label{tab:cost}
 \begin{ruledtabular}
  \begin{tabular}{lllcrlcr}
   $p_s$ & $p_f$ & $p_i$ & $L_0$ & $N_0\phantom{\frac{1}{2}}$ & $M$ &
   $N(M)$ & $N_{\rm bond}$ \\
   \hline
   0.2 & 0.6 & 0.2 & $2^3=8$ & 365$\phantom{\frac{1}{2}}$ & 9 & $185
   M$ & 3334$\phantom{\frac{1}{2}}$ \\
   0.3 & 0.4 & 0.3 & $2^2=4$ & 18$\frac{8}{9}$ & 5$\frac{2}{3}$ &
   $16\frac{2}{3} M$ & 191$\frac{2}{9}$ \\
   0.4 & 0.2 & 0.4 & $2^1=2$ & 2$\frac{1}{2}$ & 3 & $5 M$ &
   32$\frac{1}{2}$ \\
   0.5 & 0.5 & 0   & $2^2=4$ & 10$\phantom{\frac{1}{2}}$  & 5 & $6
   M$ & 62$\phantom{\frac{1}{2}}$ \\
  \end{tabular}
 \end{ruledtabular}
\end{table}

Linear clusters are not sufficient for simulating arbitrary
logic networks \cite{Nielsen2005}, and therefore it is
necessary to generate more general graph states. A variety
of techniques for making such states using probabilistic
entangling operations have been proposed, which include
linking linear clusters using independently prepared `I'
shaped clusters \cite{Sean}, using micro-clusters
\cite{nielsen04}, using redundantly encoded qubits
\cite{browne04}, or by making use of `+' shaped clusters
\cite{DuanRaussendorf}. Here, we propose a relatively
efficient method for creating vertical bonds between linear
cluster chains.

We employ a technique based on that introduced by Browne
and Rudolph \cite{browne04}, which involves four steps as
shown in Figure~\ref{fig:cluster3}.
\begin{enumerate}
 \item[a)] First, we assume that we have sufficiently long linear
   cluster chains. These can be produced efficiently in the manner
   outlined above. In order to establish the amount of resources
   needed to create a vertical bond, we will count the number of
   qubits that are utilized on average in this process, as well as the
   average number of entangling operations.
 \item[b)] Secondly, we identify the two qubits that we wish to
   entangle with a vertical bond (in Figure~\ref{fig:cluster3} the two
   left-most qubits). The qubits directly on the right of
   these qubits are then measured in the $\sigma_x$ basis. A Hadamard
   operation on the third qubit in each chain returns the overal state to a
   graph state.
 \item[c)] This will result dangling bonds, or {\em cherries}
   \cite{tom} hanging from the two qubits that are to be
   connected. This is a form of redundant encoding, and it allows us
   to apply the entangling operation to the two cherries. In case of a
   failure, the entangling operation will {\em not} break the linear
   cluster chains. It will destroy only the cherries and as a result
   both chains are shortened by two qubits. Steps (b) and
   (c) can then be repeated.
 \item[d)] When the entangling operation succeeds, we have forged a
   vertical bond between the two qubits chosen in step a). The
   vertical link is itself a chain of two qubits. These are typically
   not wanted, so we can remove one of them with a $\sigma_x$
   measurement, creating another cherry in the other qubit in the
   chain. This redundancy can be pruned, but may also be
   useful for creating additional bonds, or may even be
   useful for error correction.
\end{enumerate}

We will now estimate the cost of this procedure. Since two qubits are
burnt in each step, and we need to repeat the process $P_s^{-1}$
times, the average length of each chain that is consumed in the
bonding process is
\begin{equation}
 M = 2 P_s^{-1} + 1 = \frac{2 (1-p_i)}{p_s} + 1\; ,
\end{equation}
where the extra $+1$ counts the qubits that will establish the
vertical link. The number of entangling operations needed to make a
vertical bond is then
\begin{equation}
 N_{\rm bond} = 2 N(M) + P_s^{-1} = 2 N(M) + \frac{(1-p_i)}{p_s}\; ,
\end{equation}
where the extra $P_s^{-1}$ takes into account the number of entangling
operations that are needed to link the cherries together into a
vertical bond. In Table~\ref{tab:cost}, we calculated the
number of entangling operations that are needed to forge a vertical
bond, given several specific values for the success, failure, and
insurance probabilities. 

\section{Conclusions } \label{conc}

\noindent
We analysed a hybrid architecture for quantum computing using
stationary and flying qubits, which is based on our earlier work \cite{moonlight,Sean}, in detail. 
It was shown that this new approach solves some of the most pressing
problems that arise in non-hybrid architectures. Our system is
scalable, even with non-ideal components, and more importantly, it
uses no direct qubit-qubit interactions. This means that the qubits
will be subject to less decoherence and fewer control errors. When realistic
photo-detectors are used, photon loss will affect only the efficiency
of the scheme. Furthermore, our system relies on components that have
been demonstrated in experiment, and is largely implementation
independent. Despite the no-go
theorem for optical Bell-state measurements, it is in principle 
possible to implement a deterministic gate between distant
qubits. 

However, when losses are taken into account, the gate becomes
necessarily probabilistic. In order to achieve robustness against
general decoherence and to guarantee high fidelities, we showed how to construct cluster- or
graph-states using the two-qubit gate. Our entangling operation, which produces the bonds in the
graph states, is not limited to physically adjacent matter qubits. As
a consequence, no extensive swapping operations need to be taken into
account in the production of nontrivial graph states. This
architecture for quantum computation is inherently distributed, and
hence can be used for integrated quantum computation and communication
purposes. \\[0.5cm]

{\em Acknowledgment.} YLL, AB and LCK thank H.J. Briegel,
A. Browaeys, D.E. Browne, T. Durt, A.K. Ekert, P. Grangier, M. Jones, and
P.L. Knight for stimulating discussions. PK and SDB thank S. Benjamin,
J. Eisert, B. Lovett, M.A. Nielsen, T. Stace for fruitful discussions
and particularly D. Browne for giving them a deeper insight into
cluster state generation. YLL acknowledges funding from the DSO
National Laboratories in Singapore and AB acknowledges support from
the Royal Society and the GCHQ. This work was supported in part by the
European Union (RAMBOQ, QGATES) and the UK Engineering and Physical
Sciences Research Council (QIP IRC).


\begin{thebibliography}{99}
\bibitem{endnote}
This gate can be easily accomplished by applying a strongly detuned laser field for a certain time $t$.
\end{thebibliography}
\end{document}